\documentclass[%
reprint,
superscriptaddress,
amsmath,amssymb,
aps,
prapplied,
]{revtex4-2}

\usepackage{graphicx}
\usepackage{dcolumn}
\usepackage{bm}
\usepackage{color}
\usepackage{physics}
\usepackage[separate-uncertainty]{siunitx}
\AtBeginDocument{\RenewCommandCopy\qty\SI}

\newcommand*\diff{\mathop{}\!\mathrm{d}}
\begin{document}

\title{Focusing Surface-Acoustic-Wave Resonators on Thin-Film Lithium Niobate with Transverse-Mode Suppression}

\author{Ryo~Sasaki}
\affiliation{RIKEN Center for Quantum Computing (RQC), Wako, Saitama 351-0198, Japan}

\author{Ryusuke~Hisatomi}
\affiliation{Institute for Chemical Research, Kyoto University, Uji, Kyoto 611-0011, Japan}
\affiliation{Center for Spintronics Research Network, Institute for Chemical Research, Kyoto University, Uji, Kyoto 611-0011, Japan}

\author{Rekishu~Yamazaki}
\affiliation{International Christian University, Mitaka-shi, Tokyo 181-8585, Japan}

\author{Yuya~Yamaguchi}
\affiliation{National Institute of Information and Communication Technology, Koganei-shi, Tokyo 184-8795, Japan}

\author{Yasunobu~Nakamura}
\affiliation{RIKEN Center for Quantum Computing (RQC), Wako, Saitama 351-0198, Japan}
\affiliation{Department of Applied Physics, Graduate School of Engineering, The University of Tokyo, Bunkyo-ku, Tokyo 113-8656, Japan}

\author{Atsushi~Noguchi}
\affiliation{RIKEN Center for Quantum Computing (RQC), Wako, Saitama 351-0198, Japan}
\affiliation{Komaba Institute for Science (KIS), The University of Tokyo, Meguro-ku, Tokyo 153-8902, Japan}
\affiliation{Inamori Research Institute for Science (InaRIS), Organization, Kyoto-shi, Kyoto 600-8411, Japan}

\date{\today}

\begin{abstract}
Surface-acoustic-wave (SAW) resonators are a promising platform for constructing hybrid quantum systems, where confined acoustic waves enable strong interaction with various physical systems. Focusing SAW resonators, reducing mode volume while suppressing diffraction losses, have recently been investigated for application in such hybrid systems. However, the resonator leads to additional transverse-mode resonances, which cause undesired responses. In this work, we develop focusing SAW resonators on a thin-film lithium niobate on sapphire. A film thinner than the SAW wavelength allows a highly confined acoustic-wave mode to be localized on the substrate surface. By using contoured electrodes following a two-dimensional Gaussian beam shape, we make the SAW mode focused to nearly a diffraction-limited and confirm it via optical imaging. Furthermore, by engineering the spatial mode overlap of the interdigital transducer electrodes, we suppress the excitation of higher-order transverse modes.

\end{abstract}

\maketitle

\section{Introduction}
Hybrid quantum systems, where different physical systems are coherently coupled to each other, have been actively studied in recent years~\cite{kurizki2015}. Acoustic waves are one of the promising candidates for mediating interactions between different physical systems, owing to their slow velocity and short wavelength compared to electromagnetic waves~\cite{chu2020}. Small mode volume of acoustic waves can enhance the interaction strength with other physical degrees of freedom, which is advantageous for quantum applications.

Surface-acoustic-wave (SAW) modes are acoustic wave modes localized on the surface of a medium. They can be electrically excited on a piezoelectric substrate with an interdigital transducer~(IDT). The energy density of the SAW is highly concentrated near the surface~\cite{morgan2010surface}, which enables a strong interaction with other physical systems located on the surface. The mode confinement in the propagation direction is enabled by using Bragg mirrors as reflectors and making a SAW resonator~\cite{datta1986surface,morgan2010surface}. The SAW resonators are a suitable platform for achieving enhanced coupling between acoustic wave and other physical systems~\cite{schuetz2015}, and several hybrid systems have been studied, such as with superconducting qubits~\cite{Manenti2017, Noguchi2017, Satzinger2018, Moores2018, chou2025}, spins in solid~\cite{Whiteley2019}, electrons in quantum dots~\cite{Imany2022, DeCrescent2022, Wang2024}, electrons in two-dimensional materials~\cite{Fang2023}, magnon in ferromagnetic materials~\cite{hatanaka2022,hwang2024,kunstle2025}, and photons in optical circuits~\cite{Sarabalis2020, Okada2021}.

To further increase the coupling strength, it is effective to reduce the SAW mode volume in the planar dimension. However, simply reducing the beam width of a SAW mode causes diffraction loss, degrading the SAW-resonator quality~\cite{Aref2016}. In order to reduce the mode volume while suppressing diffraction loss, focusing SAW resonators, analogous to optical resonators with concave mirrors, have been studied~\cite{Whiteley2019, Msall2020, DeCrescent2022}. Designing focusing SAW resonators, however, remains challenging, mainly due to the anisotropy associated with piezoelectric materials. More resonance peaks than the fundamental ones can appear as the curved mirrors support various higher-order modes,  which complicates understanding the resonator properties. 

\begin{figure*}
    \centering
    \includegraphics[width = 0.8\paperwidth]{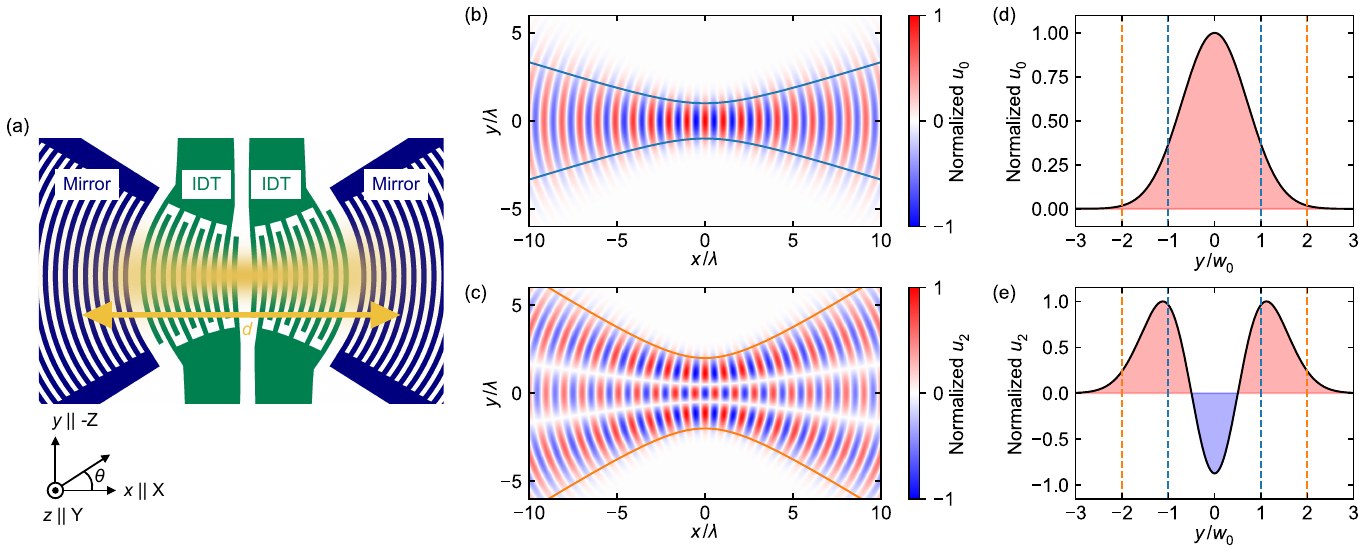}
    \caption{
        (a)~Schematic image of a 2-port focusing SAW resonator.
        The yellow arrow indicates the effective mirror distance $d$ of two Bragg mirrors.
        The coordinate axes, $\{x,y,z\}$,  and corresponding crystal axes of LN, $\{\mathrm{X, Y, Z}\}$, used in this work are shown in the inset.
        (b)(c)~Normalized displacement profile of the fundamental ($l=0$) and higher-order transverse ($l=2$) modes, respectively. Blue and orange solid curves respectively depict the mode width $\pm w(x)$ and $\pm 2w(x)$, respectively, where $w_0 \equiv w(x=0)$ is set to $\lambda$. 
        (d)(e)~Cross-sections of the normalized displacement at the mode waist $x = 0$ for the two modes.
        }
    \label{fig: SAW Gaussian beam}
\end{figure*}

In this work, we develop focusing SAW resonators using a thin film of lithium niobate on a sapphire substrate. Lithium niobate~(LN) has strong piezoelectricity and optical nonlinearity, providing a platform for developing novel quantum technologies based on hybrid quantum devices with superconducting circuits~\cite{Satzinger2018,arrangoiz2019,chou2025} and quantum transducers between microwave and optical quantum signals~\cite{ Sarabalis2020, shao2019, jiang2020}. Due to the lower acoustic phase velocity in LN with respect to that in sapphire, the SAW mode is mostly confined in the LN thin film thinner than the acoustic wavelength. By using the LN thin film, we fabricated the focusing structure of IDTs and mirrors to realize in-plane mode confinement. The resonance peaks observed in resonators with IDTs and mirrors of various curvatures are well explained by the theory based on two-dimensional SAW Gaussian modes. Optical imaging allows us to assign resonance modes and observe the beam focused to wavelength scales. To mitigate higher-order transverse modes excitation induced by curved electrodes, we introduce spatial-mode-selective coupling strategy for IDT design. It can effectively suppress the coupling to higher-order transversal modes. This design can also be used to enhance the quality factor in a small-mode-volume SAW resonator. 

\section{Theory}
\subsection{SAW Gaussian mode}
The focusing SAW resonator in this work is designed based on a Gaussian beam mode, which is well-understood in paraxial analysis of the electromagnetic waves in optics~\cite{yariv2007}. In an optical resonator with two concave mirrors with the same radius of curvature, the optical resonator modes have the electric field distribution following the Gaussian mode shape with a focus at the center of the three-dimensional~(3D) resonator. In a SAW resonator, on the other hand, the acoustic-wave mode is approximately distributed in a two-dimensional~(2D) space, and a slight modification from the conventional form of the Gaussian beam is needed.  

In a two-dimensional system, the paraxial analysis leads to a Gaussian beam mode of the SAW. We consider a SAW propagating in the $x$-direction in the $xy$-plane. The displacement vector field $\bm{u}_l$ of the $l$th-mode can be assumed as $\bm{u}_l = \hat{\bm{a}} u_l(x,y)$, where $\hat{\bm{a}}$ is a unit vector representing a polarization of the acoustic wave.
The scalar field of displacement~${u_l}(x,y)$ is expressed as~\cite{mason1971},
\begin{align}\label{SAW gaussian beam}
    \begin{split}
        {u_{l}}(x,y) =~& 
         {U_{l}} \sqrt{\frac{w_0}{w(x)}}\, H_l\!\left(\frac{\sqrt{2} \,y}{w(x)}\right)
         \exp \!\left({-{\frac{y^2}{{w(x)}^2}}}\right)\\
         &\times \exp \!\left( -ikx-i\phi_l(x,y)\right),
    \end{split}
         \\
         \phi_l(x,y) =~& \frac{k y^2}{2R_\mathrm{SAW}(x,y)} - \psi^\mathrm{G}_l(x),
\end{align}
where ${U_{l}}$ defines an amplitude, $k= 2\pi/\lambda$, and $\lambda$ is the wavelength. $w(x) = w_0\sqrt{1 + \left(x/x_R\right)^2}$ is the beam radius at $x$, where $w_0$ is the beam radius at the beam waist, and $x_\mathrm{R} = \pi w_0^2/\lambda$ is the Rayleigh length. $H_l$ is the $ l$th-order Hermite polynomial to describe the displacement field perpendicular to the beam axis $x$. The Gouy phase $\psi_{l}^\mathrm{G}(x) = \left(l + 1/2\right)\arctan\!\left(x/x_\mathrm{R}\right)$ arises from the transverse component of wavevector~\cite{mason1971,Feng2001}. In a 2D system, there is only one transverse component. This fact modifies the Gouy phase from that of the Gaussian beam in 3D. In the fundamental mode, the Gouy phase of the SAW is half that of the 3D mode and is consistent with the previous studies on the SAW Gaussian beams~\cite{Msall2020, Usami2020}.

Since the acoustic phase velocity~$v_\mathrm{p}$ in LN is anisotropic, as shown in Fig.~\ref{fig: LNonSa dispersion and anisotropy}(d), the wavefront is distorted during propagation. To incorporate this effect, the radius of curvature is set to $R_\mathrm{SAW}(x,y) = R(x) v_\mathrm{g}(\theta)/v_\mathrm{g}(0)$, where $R(x) = x[1 + \left(x_\mathrm{R}/x\right)^2]$ is the radius of curvature of the beam at $x$, and $\theta = \arctan(y/x)$ is in-plane angle from the $x$-axis. The in-plane group velocity $v_\mathrm{g}(\theta)$ can be calculated from $v_\mathrm{p}(\theta)$~\cite{Msall2020, Okada2018}.

The displacement distribution of the fundamental mode~$u_0(x,y)$ and the higher-order transverse mode~$u_2(x,y)$ are shown in Figs.~\ref{fig: SAW Gaussian beam}(b) and (c), respectively. The IDT design symmetric with respect to the $x$-axis can excite only the higher-order modes with an even number of $l$. The normalized displacements at the beam waist $(x = 0)$ for $l=0$ and $l=2$ modes are shown in Figs.~\ref{fig: SAW Gaussian beam}(d)~and~(e), respectively. In this example, the beam waist~$w_0$ is set to the wavelength~$\lambda$. While the fundamental mode shows a single Gaussian peak~[Fig.~\ref{fig: SAW Gaussian beam}(d)], the $l=2$ mode has a long distribution tail, extending beyond $y = \pm 2w_0$~[Fig.~\ref{fig: SAW Gaussian beam}(e)]. For $l=2$, there are nodes near $y = \pm w_0/2$, where the sign of the displacement changes. This sign change is crucial for suppressing the excitation in the higher-order mode of the resonator, as discussed later.

The resonance frequency of the SAW resonator can be calculated by imposing the resonator round-trip distance to be an integer multiple of the acoustic wavelength. In terms of the Gaussian beam phase $\phi_l$, the resonant condition can be given as 
\begin{equation}
    \phi_l\left(\frac{d}{2},0\right) - \phi_l\left(-\frac{d}{2},0\right)  + kd = n\pi,
\end{equation}
where $d$ is the effective distance of the two mirrors considering the finite-length penetration into the Bragg mirrors~[see Fig.~\ref{fig: SAW Gaussian beam}(a)] and integer $n$ is a longitudinal mode index of the resonator. This leads to the resonance frequency of the SAW cavity: 
\begin{equation}\label{eq: resonance frequency}
    f_{n,l} = \frac{v_\mathrm{p}}{2d}\left[n + \frac{2}{\pi}\left(l + \frac{1}{2}\right)\arctan\left(\frac{d/2}{x_\mathrm{R}}\right) \right].
\end{equation}
The resonance frequency difference~$\Delta f_{l0}$ between the fundamental~($l = 0$) and higher-order transverse~($l > 0$) modes in the same mode index~$n$ is given by
\begin{equation}\label{delta frequency}
    \Delta f_{l0} = f_{n,l} - f_{n,0} = \frac{v_\mathrm{p}}{2d}\left[\frac{2l}{\pi}\arctan\left(\frac{d/2}{x_\mathrm{R}}\right)\right].
\end{equation}
Note here that $\Delta f_{l0}$ is independent of the mode index $n$ and depends on $w_0$ through $x_\mathrm{R}$. For a tight focus, where $w_0$ is small, $x_\mathrm{R}$ becomes shorter, and $\Delta f_{l0}$ increases. This is because the smaller radius of curvature arising from a tight mode waist, small $w_0$, induces a larger deviation from planar SAW resonators with sufficiently large aperture, where all the transverse modes are nearly degenerate with the fundamental mode~\cite{fisicaro2025}. The difference in resonance frequency for different $l$ originates from the difference in the Gouy phase $\phi_l^\mathrm{G}$, which vanishes in the limit of large $x_\mathrm{R}$. $\Delta f_{l0}$ also increases with $l$.

\subsection{LN on sapphire}
\begin{figure}
    \centering
    \includegraphics[width = \linewidth]{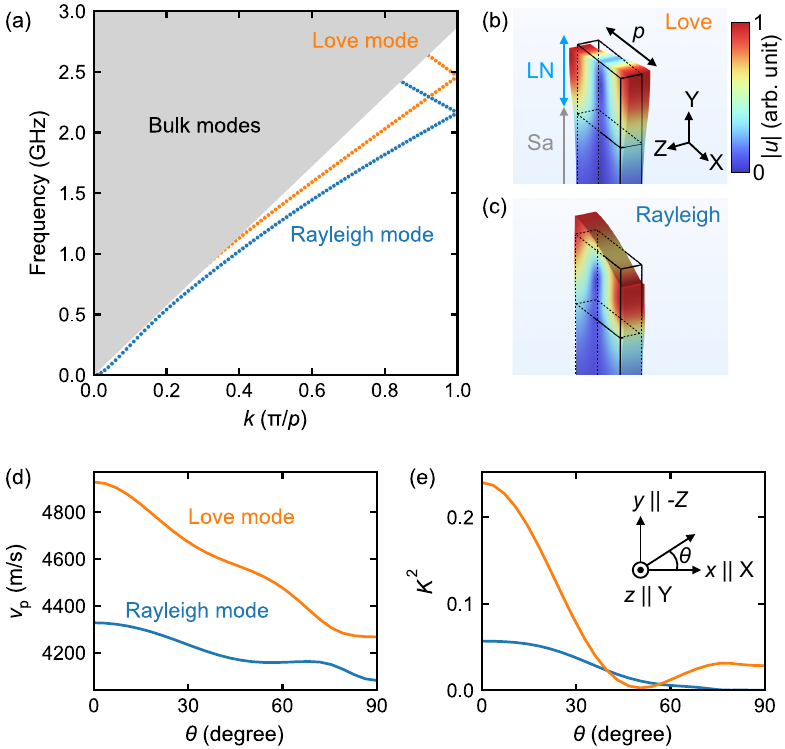}
    \caption{
            (a)~Simulated dispersion relation of the SAW modes in an LN thin film with a thickness of $\SI{0.8}{\micro\meter}$ on a sapphire substrate. The gray area shows the continuum of bulk-acoustic-wave modes in sapphire. $p$ = \SI{1}{\micro\meter} is the unit length of the model along the X-axis. The $k^2$-dependent dispersion at $k\sim0$ is an artifact due to the finite length of the model along the Y axis, which is comparable to the mode wavelength in this region.
        (b)(c)~Displacements $u$ of Love and Rayleigh modes at $k = \pi/p$, respectively. LN and Sa represent the region of lithium niobate and sapphire, respectively
        (d)(e)~Simulated crystal anisotropy of the acoustic phase velocity $v_\mathrm{p}$ and electromechanical coupling $K^2$ of the SAW modes.
        }
    \label{fig: LNonSa dispersion and anisotropy}
\end{figure}

The SAW devices are fabricated using a Y-cut LN thin film with a thickness of \SI{0.8}{\micro\meter} on a c-cut sapphire substrate~(NGK Insulators). Since the acoustic phase velocity $v_\mathrm{p}$ of LN is much slower than that of the sapphire, the SAW mode is tightly confined within the LN thin film~\cite{Safavi-Naeini2019}.  In this work, the SAW wavelength of $\lambda = \SI{2}{\micro\meter}$ longer than the film thickness is used. Figure~\ref{fig: LNonSa dispersion and anisotropy}(a) shows the acoustic-wave dispersion relation in the configuration, simulated with COMSOL~\cite{comsol}.  The wavevector $k$ is along the X-axis of LN. The anisotropy of sapphire is neglected in the simulation, as the field penetrated the substrate is minimal. Here, the simulation does not include the metal-electrode structures. The gray area in the plot shows the continuum of the bulk modes in the sapphire substrate. There are two SAW modes whose phase velocities are below the velocities of the bulk modes. The displacements of these SAW modes at $k = \pi/p$ are shown in Figs.~\ref{fig: LNonSa dispersion and anisotropy}(b) and~(c), where $p = \SI{1}{\micro\meter}$ is the model unit length along the  X-axis. Since these modes mainly have in-plane transverse and out-of-plane vertical displacements, they can be assigned to Love and Rayleigh modes, respectively. The calculation shows that both the Rayleigh and Love modes are mostly confined in the thin-film region, whose thickness is shorter than the wavelength. This is in contrast with the conventional SAW mode on a bulk substrate, where the displacement is distributed in the depth of approximately one wavelength.

Figures~\ref{fig: LNonSa dispersion and anisotropy}(d) and (e) show the simulated in-plane anisotropy of phase velocity and electromechanical coupling $K^2$ for the Y-cut LN on sapphire, respectively. Since the phase-velocity variation is symmetric across the X-axis, the power flow angle is also parallel to the X-axis, which tremendously reduces the wave leak loss from the resonator~\cite{morgan2010surface}. The electromechanical coupling $K^2$ can be calculated from the difference of simulated phase velocities between the electrically-free~($v_\mathrm{p}^\mathrm{free}$) and shorted~($v_\mathrm{p}^\mathrm{short}$) boundary conditions of the surface~\cite{morgan2010surface}: 
\begin{equation}
    K^2 = 2\frac{v_\mathrm{p}^\mathrm{free} - v_\mathrm{p}^\mathrm{short}}{v_\mathrm{p}^\mathrm{free}}.
\end{equation}
The excitation efficiency of IDTs and the reflectivity of the Bragg mirrors made of metal electrodes can be calculated from $K^2$. Because  $K^2$ of the Love mode propagating along the X-axis is 4.2 times higher than that of the Rayleigh mode, as shown in Fig.~\ref{fig: LNonSa dispersion and anisotropy}(e), the Love mode is excited more effectively~\cite{Kuznetsova2001}. The higher $K^2$ increases the reflectivity of electrodes and makes the penetration depth into the Bragg mirrors shorter, which contributes to the reduction in the resonant mode volume. For the subsequent experiment, the Love mode propagating in the X-axis is chosen to realize the focusing resonator with a small mode volume.

\section{Experiment}
\subsection{Device design}
\begin{figure}
    \centering
    \includegraphics[width = \linewidth]{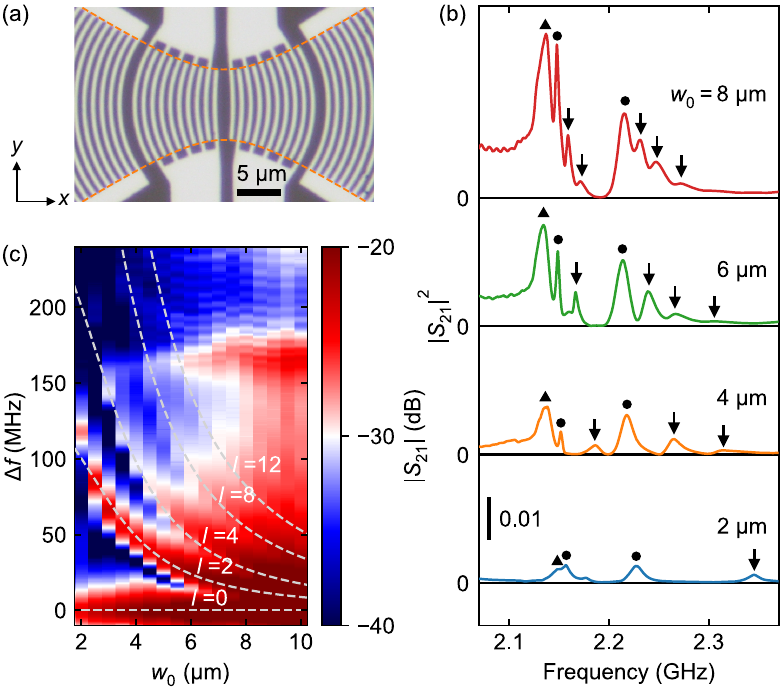}
    \caption{(a)~Optical microscope image of a fabricated device.  The orange dashed lines show the designed electrode length of~$\pm 2w(x)$.  
    (b)~Microwave transmission spectra $|S_{21}|^2$ between two IDTs. 
    The black arrows indicate the resonance peaks attributed to transverse modes.
    Circles and a triangle indicate fundamental and spurious modes, respectively.
    (c)~Shift of the resonant frequencies of the higher-order modes. The transmission spectra $|S_{21}|$ as a function of the frequency separation $\Delta f$ from the fundamental-mode frequency at around \SI{2.21}{\giga\hertz} are measured in 17 devices with different $w_0$. The white dashed lines are the frequency separations between the fundamental~($l=0$) and higher-order~($l>0$) modes $\Delta f_{l0}$, for $l$ =0, 2, 4, 8, and 12, calculated from Eq.~(\ref{delta frequency}).}
    \label{fig: non-TOE_device}
\end{figure}
The SAW-resonator design of a typical device is shown in Fig.~\ref{fig: non-TOE_device}(a). The curved mirrors and IDTs are designed to match the wavefront of the acoustic Gaussian mode. A tight mode waist size of $w_0 \equiv w(x=0) = \SI{2}{\micro\meter}$ is used for this particular device. The overlap length of the IDTs and mirrors, i.e.,  an excitation aperture, is set to $\pm 2w(x)$ as shown in Fig.~\ref{fig: SAW Gaussian beam}(c). The devices are fabricated with various beam-waist sizes $w_0$ ranging from 2 to \SI{10}{\micro\meter}. 

All the mirror and IDT electrodes are made of aluminum with \SI{50}{\nano \meter} thickness, fabricated using electron beam lithography and electron beam evaporation. The number of IDT finger pairs and mirror fingers is 5 and 200, respectively. Mirror electrodes are shunted, forming a shorted-electrodes mirror. The width and spacing of each electrode are set to $\lambda/4$ = \SI{500}{\nano \meter}, and the electrode pitches are \SI{1}{\micro\meter}. The center of the IDT electrode is positioned at the antinode of the acoustic mode to obtain a large coupling, while the inner edge of the mirror electrode is placed at the node to achieve high reflectivity for the mirror with shorted electrodes~\cite{Haydl1976}. 

\subsection{Device characterization with microwave measurement}
Figure~\ref{fig: non-TOE_device}(b) shows microwave transmission spectra between two IDTs for the devices with various $w_0$. The spectra are measured with a vector network analyzer at room temperature. In the spectra, several resonance peaks are observed in the frequency range determined by the stop band of Bragg mirrors ($\sim \SI{0.2}{\giga\hertz}$). We notice that the peaks depicted with black arrows show large shifts toward lower frequencies with increasing $w_0$, and their spacings decrease. This dependency is consistent with the transverse-mode resonance frequency in Eq.~(\ref{eq: resonance frequency}). Two peaks marked with circles at \SI{2.15}{\giga\hertz} and \SI{2.23}{\giga\hertz} show smaller shifts and are assigned to the fundamental SAW Gaussian modes with different longitudinal mode index $n$. The small shifts in the resonance frequency of the fundamental modes can be caused not only by the change in $w_0$ as in Eq.~(\ref{eq: resonance frequency}) but also by the thickness variation of the LN film depending on the device position on the chip~(see the discussion in Appendix~\ref{section: Thickness effect}). The broad peak marked with a triangle at \SI{2.14}{\giga\hertz} is attributed to a spurious acoustic resonance mode, as confirmed in the optical imaging measurement discussed in Appendix~\ref{section: imaging of spurious mode}.

To confirm that the observed sub-peaks are in fact the transverse modes, we measured the transmission spectra of devices with various waist sizes $w_0$. Figure~\ref{fig: non-TOE_device}(c) shows the transmission spectra as a function of frequency difference $\Delta f$ from the second fundamental modes at around \SI{2.21}{\giga\hertz} for various waist sizes $w_0$ ranging from 2 to \SI{10}{\micro\meter} with a step of \SI{0.5}{\micro\meter}. There are several branches of the resonance peaks, which show clear dependence on $w_0$. We overlay the lines indicating the frequency spacing $\Delta f_{l0}$ for $l$ = 0, 2, 4, 8, and 12 calculated from Eq.~(\ref{delta frequency}) with the mirror distance $d$ set to the effective cavity length~(\SI{33.6}{\micro\meter}) calculated from the free spectral range of the fundamental modes. The observed $w_0$-dependence of the $\Delta f$ is well reproduced from the theoretical calculation of~$\Delta f_{l0}$.

\subsection{Optical imaging of SAW resonance modes}

\begin{figure*}
    \centering
    \includegraphics[width = 0.8\paperwidth]{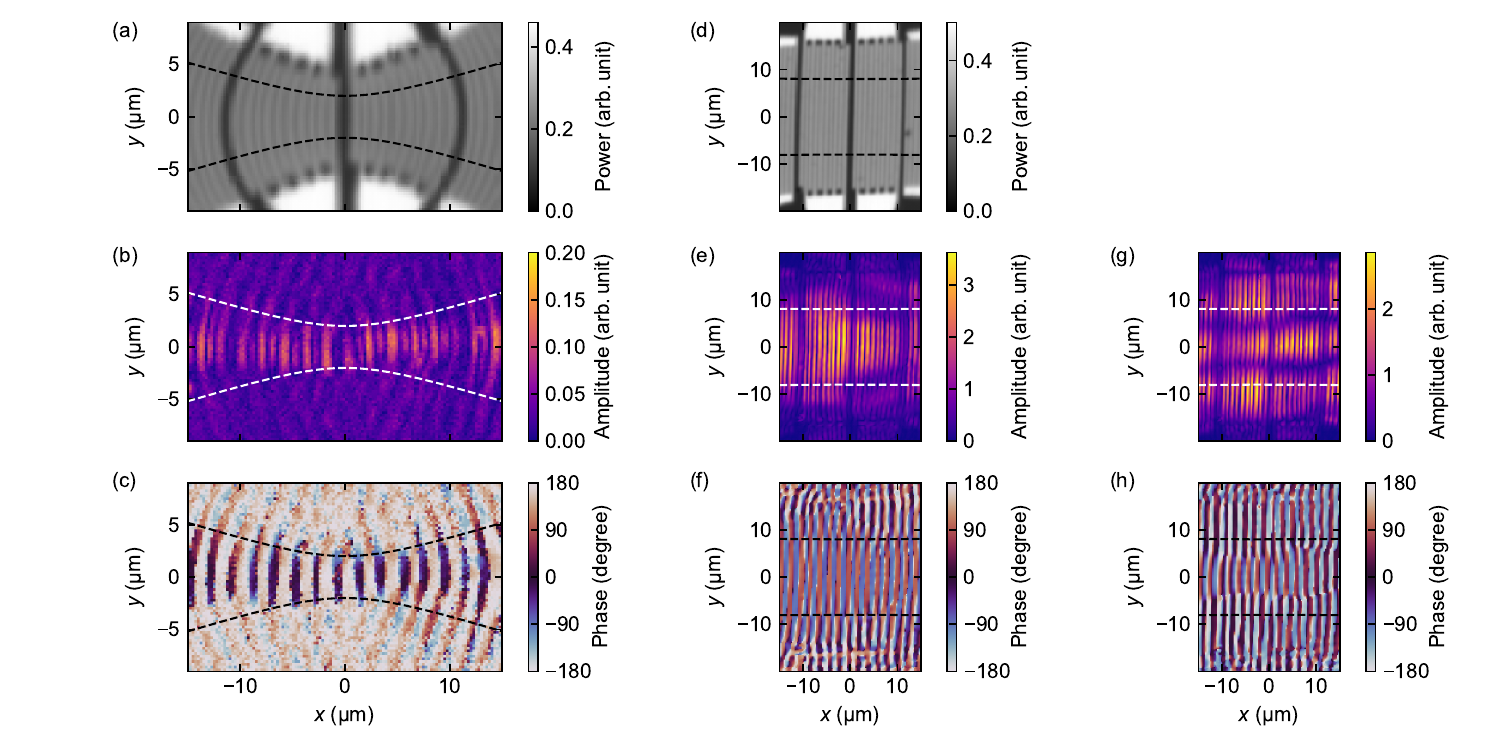}
    \caption{Optical imaging of focusing SAW resonators. 
    (a)~Optical power measured with a power meter for the SAW resonator with $w_0 = \SI{2}{\micro\meter}$ at the resonance frequency of \SI{2.159}{\giga\hertz}. (b)~Amplitude, and (c)~phase of the optical signal measured with a lock-in amplifier.
    (d)~Optical power, (e)~amplitude, and (f)~phase for the SAW resonator with $w_0 = \SI{8}{\micro\meter}$ at the resonance frequency of \SI{2.148}{\giga\hertz}. 
    (g)~Amplitude and (h)~phase at \SI{2.159}{\giga\hertz}, corresponding to the $l=2$ transverse mode. The scanning areas are the same as in (d)--(f).
    Dashed lines in (a)--(c) and (d)--(h) depict the mode width $\pm w(x)$ for $w_0 = \SI{2}{\micro\meter}$ and $\SI{8}{\micro\meter}$, respectively}
    \label{fig: imaging results}
\end{figure*} 

To confirm the mode assignment and evaluate the focusing effects, we utilized optical imaging of the SAW resonance mode. The optical imaging of the SAW mode is based on the optical-path modulation of a laser beam reflected from the SAW-induced surface tilt~\cite{Taga2021, Hisatomi2023}. Details of the imaging system are described in Appendix~\ref{section: imaging system}.

Figure~\ref{fig: imaging results}(a) shows the signal measured with the power meter, and Figs.~\ref{fig: imaging results}(b) and (c) show the amplitude and phase of the optical signal caused by the oscillations of the SAW resonator with $w_0$ = \SI{2}{\micro\meter}, respectively. The frequency of the microwave drive is set to the resonance frequency of the first fundamental mode at \SI{2.159}{\giga\hertz}. Since the aluminum electrodes have larger optical reflectivity than LN, the DC signal from the power meter is large at the surface of the IDTs and mirrors, and the image reproduces the actual device structure as shown in Fig.~\ref{fig: non-TOE_device}(a). Both the amplitude and phase of the AC signals in Figs.~\ref{fig: imaging results}(b) and (c) show a single central lobe along the $x$-axis, and no node is observed in the direction along the $y$-axis, which is consistent with the fundamental ($l = 0$) mode of the SAW Gaussian beam. The phase of the AC signal shows a periodic oscillation along the $x$-axis with a period of about \SI{2}{\micro\meter}, which is consistent with the periodicity of the SAW displacement shown in Fig.~\ref{fig: SAW Gaussian beam}(b). 

Figures~\ref{fig: imaging results}(d)--(h) show the imaging results of the device with a larger mode waist ($w_0=\SI{8}{\micro\meter}$). Figures~\ref{fig: imaging results}(e)~and~(f) are the amplitude and phase of the AC signal, respectively, with the microwave drive frequency set to the fundamental mode at \SI{2.148}{\giga\hertz}. The imaging result shows that the mode profile has a single central lobe in the $y$-direction, characteristic of the fundamental Gaussian mode. The asymmetry in the $x$-direction of the amplitude image may originate from the one-port excitation during the optical measurement, where the SAW is excited from only the right-side IDT, and the other port is floating. Figures~\ref{fig: imaging results}(g)~and~(h) show the amplitude and phase with the microwave drive frequency set to $l=2$ transverse mode at \SI{2.159}{\giga\hertz}. In the amplitude image, there are two nodes around $y \sim \pm \SI{4}{\micro\meter} = w_0/2$ line, and the phase shows steep changes along the $y$ direction on this line. This is consistent with the displacement distribution of the $l = 2$ mode shown in Fig.~\ref{fig: SAW Gaussian beam}(c) and confirms the mode assignment in the microwave measurement.

\begin{figure}
    \centering
    \includegraphics[width = 1.0\linewidth]{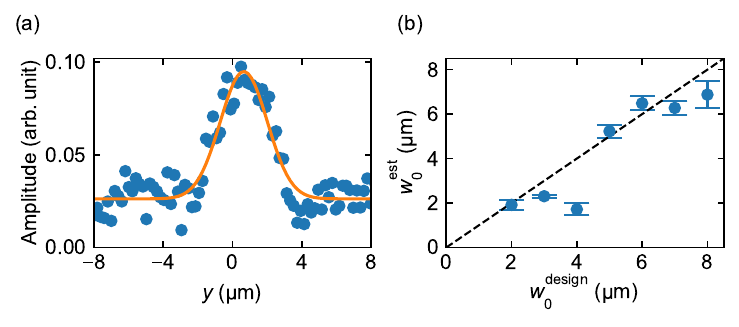}
    \caption{
    (a)~Amplitude profile along the $y$-axis at the mode waist of the device with $w_0 =$ \SI{2}{\micro \meter}. The solid curve shows the fit to a Gaussian function. 
    (b)~Estimated half mode waist $w_0^\mathrm{est}$ vs.\ the designed value  $w_0^\mathrm{design}$ of the focused SAW mode at the waist. Error bars show the fitting errors. The dashed line shows the design value.
    }
    \label{fig: w0 fitting}
\end{figure}

The waist $w_0$ of the fundamental mode at around \SI{2.159}{\giga\hertz} is estimated from the fit to the amplitude profile near $x=0$. Figure~\ref{fig: w0 fitting}(a) shows the amplitude along the $y$-axis of the device with $w_0 = \SI{2}{\micro \meter}$. The fitting result shows the waist $w_0$ =\SI{1.9 \pm 0.1}{\micro \meter}, which agrees with the design. Figure~\ref{fig: w0 fitting}(b) shows a comparison between the observed waist size fitted from the imaging results and the designed mode waist for devices with various $w_0$. We observe good agreement between them. The deviation from the design, especially for the $w_0$ = \SI{4}{\micro \meter} device, can be attributed to interference with the other mode. The frequency of the $l=2$ mode of lower mode index $n$ is close to the fundamental mode of the $w_0$ = \SI{4}{\micro \meter} device, resulting in the wavefront pattern with a feature of $l=2$ mode.
Fitting this wavefront with a single-peak Gaussian effectively reduces the estimated mode waist.

\subsection{Transverse-overlap-engineered IDTs for suppressing higher-order transverse modes}
Although the above results demonstrate successful beam focusing, it is desirable to construct a SAW resonator that operates in a single mode. To suppress the excitation of higher-order transverse modes, we optimize the overlap between the SAW resonator mode and the IDT electrodes. The conversion efficiency $E_l(L)$ of the $l$th order mode, defined as the ratio of the power in the IDT electrodes to the total power of the mode, can be estimated by~\cite{waldron1972, Yamamoto1998}: 
\begin{equation}
    E_l(L) = \frac{\left|\int_{-L}^{L}{u_l(0,y)} \diff y\right|^2}{2L \int_{-\infty}^{\infty}{ u_l(0,y)}^2 \diff y} \times \eta,
\end{equation}
where $\pm L$ defines the length of integration along the $y$-axis of the Gaussian modes shown in Figs.~\ref{fig: SAW Gaussian beam}(d) and~(e).  Since the IDT electrodes are aligned along the Gaussian wavefront in our device, we can evaluate the overlap integral at $x=0$ as a representative cross-section. For the coupling estimation, $L$ is determined from the overlap length of the IDT electrodes. The devices shown in Fig.~\ref{fig: non-TOE_device}(a) have an overlap length of $L = 2w_0$. The factor $\eta$ accounts for contributions other than the transverse overlap, including the frequency response of the IDT, the electromechanical coupling coefficient, and the overlap between the IDT and the resonance mode along the $x$-axis. In the present estimation, $\eta$ is treated as transverse-mode-independent. 

\begin{figure}
    \centering
    \includegraphics[width = 1.0\linewidth]{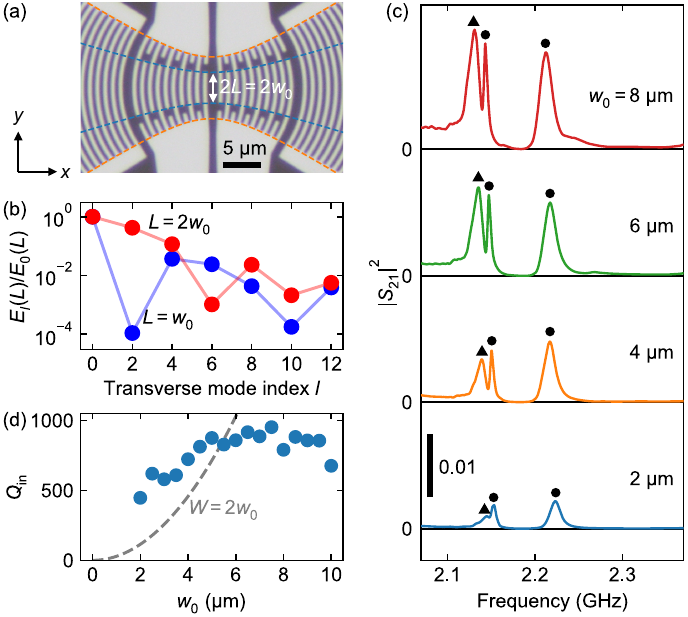}
    \caption{
    (a)~Optical microscope image of the focusing SAW resonator with TOE IDT electrodes.  The dashed blue and orange lines show $\pm w(x)$ and $\pm 2 w(x)$, respectively.
    (b)~Conversion efficiency $E_l(L)$ of the $l$th-order modes normalized by that of the fundamental mode, $E_0(L)$,  for $L = 2w_0$ and $w_0$. 
    (c)~Microwave transmission spectra $|S_{21}|^2$ of the devices with TOE IDTs. 
    Circles and triangles indicate nearly $w_0$-independent fundamental and spurious modes.
    (d)~Internal quality factor $Q_\mathrm{in}$ of the fundamental modes around \SI{2.15}{\giga\hertz} for the devices with TOE IDTs.
    Dashed lines show the diffraction-limited quality factor $Q_\mathrm{d}$ for planar SAW resonators with full beam width $W = 2w_0$.}
    \label{fig: TOE_device}
\end{figure}

Normalized conversion efficiencies $E_l(L)/E_0(L)$ for $L = 2w_0$ and $w_0$ are plotted in Fig.~\ref{fig: TOE_device}(b) as a function of $l$. For $L = 2w_0$, the coupling to the $l=2$ modes is almost the same as that of the fundamental mode. On the other hand, for $L = w_0$, the coupling to the $l=2$ mode is effectively suppressed. This suppression can be understood from the sign change of the displacement around $y = \pm w_0/2$ in $l=2$ mode, as shown in Fig.~\ref{fig: SAW Gaussian beam}(e). This sign change leads to destructive cancellation of coupling to the IDT electric potential and results in a small conversion efficiency. The coupling efficiency in the case of $L = 2w_0$ for higher $l$ modes is consistent with the observed spectra. The transverse modes $ l=2,4,8,$ and~12 show a larger efficiency than other modes, and corresponding peaks are observed as shown in Fig.~\ref{fig: non-TOE_device}(c).

The conversion-efficiency estimation above indicates that the coupling to the higher-order transverse modes can be suppressed by engineering the overlap length of the IDT electrodes. This was originally developed for the planar SAW resonators~\cite{Yamamoto1998}. To exhibit the suppression effect, we designed and fabricated IDTs with effective aperture truncation as shown in Fig.~\ref{fig: TOE_device}(a). The IDT electrodes have a gap for each finger to restrict the electrode overlap within $\pm w_0$. For convenience, we here refer to the IDTs as transverse-overlap-engineered (TOE) IDTs. Figure~\ref{fig: TOE_device}(c) shows the microwave transmission spectra of the devices with TOE IDTs for various $w_0$. In the spectra, two peaks from the fundamental modes and a broad peak from the spurious mode are still observed, while multiple transverse modes are clearly suppressed. The difference in the linewidth between the two fundamental mode peaks at higher and lower frequencies originates from the different coupling with the external ports~\cite{andersson2022}. Resonance modes with different longitudinal mode indices $n$ cause different mode matching with IDTs. Whether the center of the device is an antinode or a node alternates depending on whether $n$ is even or odd. This results in differences in coupling with the external ports and quality factors.

Suppressing the higher-order modes enables the determination of the quality factor for the fundamental mode more accurately. Figure~\ref{fig: TOE_device}(d) shows the internal quality factors $Q_\mathrm{in}$ of the fundamental modes around \SI{2.15}{\giga\hertz}. For tightly focused devices ($w_0 <$  \SI{5}{\micro \meter}), the quality factor shows $w_0$ dependence, which indicates that the diffraction loss is still dominant in this regime. As a reference, dashed line in Fig.~\ref{fig: TOE_device}(d) shows a theoretical diffraction-limited quality factor $Q_\mathrm{d}$ for planar SAW resonators, which is given by~\cite{li1974, Aref2016},
\begin{equation}
    Q_\mathrm{d} = \frac{5\pi}{\left|1 + \gamma \right|}\left(\frac{W}{\lambda}\right)^2,
\end{equation}
where $\gamma = -0.45$ is the diffraction parameter and calculated from the phase-velocity anisotropy shown in Fig.~\ref{fig: LNonSa dispersion and anisotropy}(d). $W$ is a full beam width of the planar SAW mode, which is typically determined by the aperture size of the planar IDT and mirrors. The focusing resonators show lower diffraction loss than the theoretical loss in the planar resonators with  $W =2w_0$. This indicates that the focusing resonator is more effective than the planar resonator in suppressing diffraction loss.

For loosely focused devices with $w_0 \geq$ \SI{5}{\micro \meter}, the quality factor is almost constant. The possible dominant loss channel in this regime is the scattering to the bulk mode from Bragg mirrors. Due to the high electromechanical coupling $K^2$ of the LN thin film of this orientation shown in Fig.~\ref{fig: LNonSa dispersion and anisotropy}(e), the Bragg mirror has a large acoustic impedance mismatch, which possibly generates a large scattering from the surface mode to the bulk modes while reducing the wave penetration into the Bragg mirrors. The coupling to the bulk modes can be reduced by using LN orientation with weaker electromechanical coupling or by using shallow groove mirrors~\cite{schuetz2015}.

\section{Conclusion}
We designed and fabricated SAW resonators with a focusing Gaussian mode in thin-film lithium niobate on sapphire. By using a thin film with a slower acoustic velocity than in the substrate, we confined the acoustic waves within the film thinner than the wavelength. Through microwave spectroscopy, we observed that the fundamental mode and higher-order transverse modes had the frequency spacing expected from the theoretical calculation. The optical imaging confirmed that the excited acoustic modes had a displacement distribution following the beam modes for the fundamental and higher-order transverse modes. The beam waist of the fundamental mode was focused to a wavelength scale as designed. By optimizing the overlap length of the IDT electrodes, we effectively suppressed the coupling to the higher-order modes while retaining the excitation of fundamental mode. Our result can be applied to designing focusing SAW resonators with a small mode volume and improving the performance of the hybrid system based on acoustic waves.

\begin{acknowledgments}
The authors acknowledge the Superconducting Quantum Electronics Research Team, the Superconducting Quantum Computing System Research Unit, and the Semiconductor Science Research Support Team for their support in device fabrication at the RIKEN Nanoscience Joint Laboratory. This work was supported in part by JST Moonshot R\&D Program (Grant No.~JPMJMS226C), JST~ERATO (Grant No.~JPMJER1601), JSPS~KAKENHI (Grant No.~JP22K13987), JST~PRESTO (Grant No.~JPMJPR2355).
R.S. was supported by the Grant-in-Aid for JSPS Research Fellow (No.~20J00325).

\end{acknowledgments}

\appendix

\section{Resonance-frequency shift due to the non-uniformity of film thickness}\label{section: Thickness effect}
\begin{figure}
    \centering
    \includegraphics[width = \linewidth]{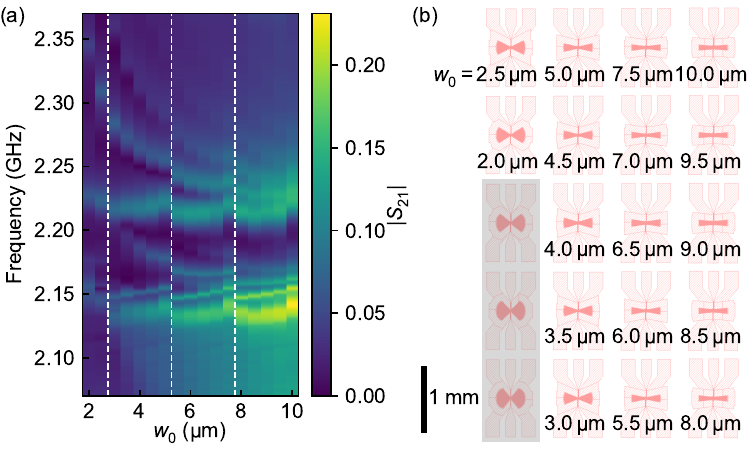}
    \caption{
    (a)~Microwave transmission spectra $|S_{21}|$ of 17 devices with different $w_0$ from 2 to $\SI{10}{\micro\meter}$. Vertical dashed lines depict the periodicity of the resonance-frequency shifts.
    (b)~Device allocations on a chip. Each object represents a 2-port focusing SAW resonator with a different $w_0$. The devices colored in the gray area did not exhibit resonance due to shorting of the electrodes.
    }
    \label{fig: raw spectrum and device array}
\end{figure}
In addition to the shift induced by the $w_0$ difference as expected in Eq.~(\ref{eq: resonance frequency}), non-uniformity of the LN film thickness also causes the device-dependent shift of the fundamental modes. Figure~\ref{fig: raw spectrum and device array}(a) shows transmission spectra of full-aperture SAW resonators with different $w_0$. The spectra show periodic resonance frequency shifts for every five devices. This periodicity corresponds to the positions of the devices on the chip as shown in Fig.~\ref{fig: raw spectrum and device array}(b). It indicates that the shift is caused by a SAW velocity change due to the LN thin-film non-uniformity. The resonance-frequency shift observed in the devices between $w_0 = \SI{7.5}{\micro\meter}$ and $\SI{8.0}{\micro\meter}$ is about \SI{9}{\mega\hertz}, corresponding to the LN thickness difference of about $\SI{20}{\nano\meter}$ calculated by the simulation. According to the data provided by the vendor, the 4-inch wafer used for device fabrication has a non-uniformity of $\SI{100}{\nano\meter}$ across the wafer. Therefore, the thickness variance in the chip is reasonable to explain the frequency shift.

\section{Measurement setup for SAW imaging}\label{section: imaging system}
\begin{figure}
    \centering
    \includegraphics[width = \linewidth]{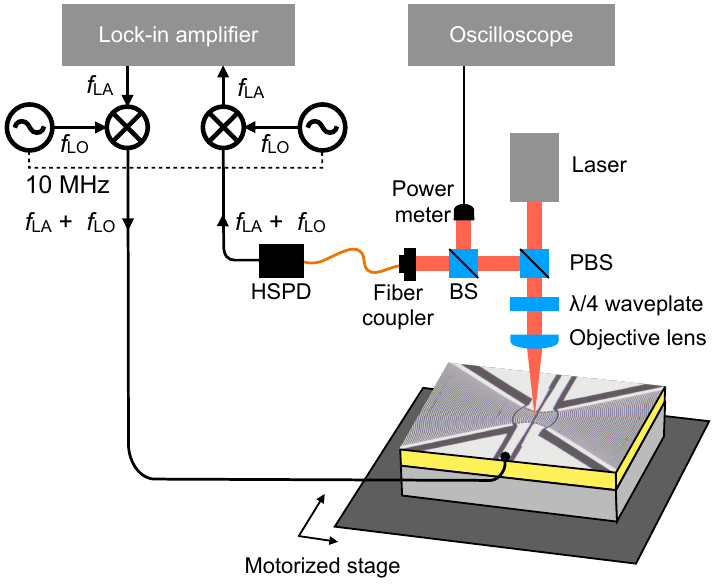}
    \caption{Schematic of the optical imaging system. The laser beam is focused on the device surface by the objective lens, and the reflected light is collected with the fiber coupler. 
    A polarized beam splitter (PBS) and a $\lambda/4$ waveplate are used to circulate the optical path.
    The AC signal is measured with a high-speed photodetector~(HSPD) and a lock-in amplifier. The DC power is measured with a power meter and an oscilloscope. 
    Black dashed line represents a 10-MHz reference signal connected with two local oscillators.}
    \label{fig: imaging system}
\end{figure}

Figure~\ref{fig: imaging system} shows the schematic of the optical imaging system. The optical imaging of SAW modes is based on the optical-path modulation of the reflection from the SAW device surface caused by the oscillating surface tilt~\cite{Taga2021, Hisatomi2023}. The input laser with a wavelength of \SI{1064}{\nano\meter} is focused on the device surface with a high-numerical-aperture (NA = 0.70) objective lens. The SAW device is mounted on the motorized stage, and the laser reflection from the device is measured while scanning the focal position of the light beam. 

The finite tilt of the surface induced by the SAW excitations causes a change of reflected direction from the normal. The microwave input to the device at the resonance frequency causes the oscillations of the surface tilt, which generate the path oscillations of the reflected laser. These oscillations can be detected by a high-speed photodetector (HSPD) via the fiber-coupling modulation. The position of the reflected laser beam guided to the fiber coupler is slightly shifted from the fiber center to increase the signal-to-noise ratio~\cite{Taga2021}. 

The microwave signal from the lock-in amplifier at frequency~$f_\mathrm{LA}$ is up-converted by mixing with the local oscillator~$f_\mathrm{LO}$, and the output microwave signal from the HSPD is down-converted by using another local oscillator at $f_\mathrm{LO}$. Then the signal is demodulated with a lock-in amplifier to obtain the amplitude and phase of the optical modulation. The amplitude is proportional to the slope of the surface tilt, which is also proportional to the amplitude of the SAW displacement.

The reflected power is also monitored with a power meter by splitting the light using a beam splitter~(BS) before the fiber coupler. The power is dependent on the reflectivity of the surface material on which the laser is focused. The DC signal from the power meter and HSPD are simultaneously measured with an oscilloscope.

The imaging resolution is determined by the focal spot size of the laser. In this work, the spot size estimated from the knife-edge method is about \SI{0.7}{\micro \meter} (not shown), which is reaching the diffraction limit of the laser with the wavelength of \SI{1064}{\nano\meter}. Although it is slightly larger than a quarter of the SAW wavelength (\SI{0.5}{\micro\meter}) and causes a decrease in the resolution and signal-to-noise ratio, it is still sufficient to detect the electrode position and the modal distribution as shown in the main text.

\section{Imaging the spurious mode}\label{section: imaging of spurious mode}

\begin{figure}[h]
    \centering
    \includegraphics[width = \linewidth]{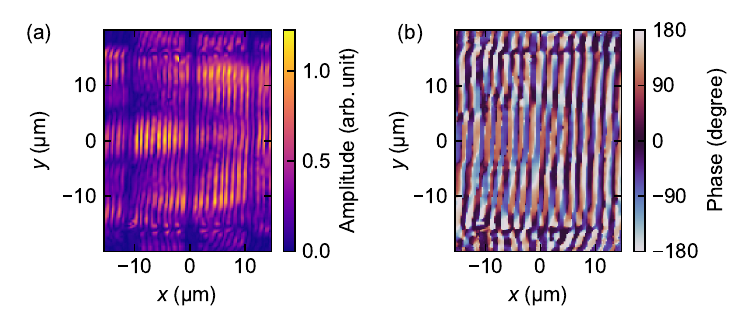}
    \caption{
    Optical imaging of the full-aperture SAW resonator with $w_0=8$~$\mu$m at the drive frequency of 2.137~GHz. The device used is the same as in Fig.~4(d).
    }
    \label{fig: spurious mode}
\end{figure}

In Fig.~\ref{fig: non-TOE_device}(b), there are two resonance peaks around \SI{2.14}{\giga\hertz} and \SI{2.15}{\giga\hertz} indicated by triangles and circles, respectively. Since these peaks show smaller frequency shifts for different $w_0$, they can be attributed to the fundamental modes of the Gaussian SAW or spurious modes, not to the higher-order transverse modes of the SAW Gaussian mode.

In the main text, we assigned the peaks marked with circles and triangles to the fundamental and spurious modes, respectively. We confirmed this assignment by imaging the resonance mode marked with a triangle. Figure~\ref{fig: spurious mode} shows the optical imaging results of the focused SAW resonator with $w_0$ = \SI{8}{\micro\meter} at the frequency of \SI{2.137}{\giga\hertz}. The imaging area is the same as in Fig.~\ref{fig: imaging results}(d). The image shows the amplitude and phase oscillations originating from the acoustic resonance. However, the observed mode is widely distributed within the resonator, exhibiting a non-single-peak distribution along the $y$-axis. This differs from the single-peak distribution shown in Fig.~\ref{fig: imaging results}(e). These results indicate that the mode is not the fundamental mode of the SAW Gaussian beam. Since similar broadly distributed mode profiles are observed for the peaks marked with triangles in other devices with different $w_0$, we concluded that the modes marked with circles and triangles are the fundamental and spurious modes, respectively. One of the possible origins of these spurious modes is the resonance caused by the unintended reflection of IDTs. Because of the strong electromechanical coupling $K^2$, IDTs also work as mirrors. This effect can induce the resonance mode distributed between IDTs and Bragg mirrors. The resonance can be mitigated by setting the distance between the IDT and the mirror to be the same as the electrode pitches of the IDT and the mirror.
Identifying the origin of the spurious mode requires measurements on additional devices with various distances between the IDT and the mirror, which is beyond the focus of this study. 

\bibliography{sawfocusing}

\end{document}